\documentclass[reprint, nofootinbib, amsmath,amssymb, aps, pra, 10pt]{revtex4-2}

\usepackage{xurl}
\usepackage{svg}
\usepackage[capitalise]{cleveref}
\usepackage{pgfplots}
\usepgfplotslibrary{groupplots}
\usepgfplotslibrary{fillbetween}
\pgfplotsset{compat=newest}

\newcommand{\vtheta}[0]{\boldsymbol{\theta}}
\newcommand{\hvtheta}[0]{\boldsymbol{\widehat{\theta}}}

\usetikzlibrary{external}
\begin{document}
	\title{Application of machine learning to experimental design in quantum mechanics}
	
	\author{Federico Belliardo}
	\affiliation{NEST, Scuola Normale Superiore, I-56126 Pisa, Italy}
	\author{Fabio Zoratti}
	\affiliation{Scuola Normale Superiore, I-56126 Pisa, Italy}
	\author{Vittorio Giovannetti}
	\affiliation{NEST, Scuola Normale Superiore and Istituto Nanoscienze-CNR, I-56126 Pisa, Italy}

	\begin{abstract}
	The recent advances in machine learning hold great promise for the fields of quantum sensing and metrology. With the help of reinforcement learning, we can tame the complexity of quantum systems and solve the problem of optimal experimental design. Reinforcement learning is a powerful model-free technique that allows an agent, typically a neural network, to learn the best strategy to reach a certain goal in a completely a priori unknown environment. However, in general, we know something about the quantum system with which the agent is interacting, at least that it follows the rules of quantum mechanics. In quantum metrology, we typically have a model for the system, and only some parameters of the evolution or the initial state are unknown. We present here a general machine learning technique that can optimize the precision of quantum sensors, exploiting the knowledge we have on the system through model-aware reinforcement learning. This framework has been implemented in the Python package qsensoropt, which is able to optimize a broad class of problems found in quantum metrology and quantum parameter estimation. The agent learns an optimal adaptive strategy that, based on previous outcomes, decides the next measurements to perform. This approach works for both Bayesian estimation and frequentist estimation. The user is required to implement the physics of the system to be studied and state which parameters in the experiment are controllable and which are unknown. The functions of the library then allow the training of the agent to optimize the precision of the sensor in a Monte Carlo simulation of the experiment. We have explored some applications of this technique to NV centers and photonic circuits. So far, we have been able to certify better results than the current state-of-the-art controls for many cases. The machine learning technique developed here can be applied in all scenarios where the quantum system is well-characterized and relatively simple and small. In these cases, we can extract every last bit of information from a quantum sensor by appropriately controlling it with a trained neural network. The qsensoropt software is available on PyPI and can be installed with pip.
	\end{abstract}
	
	\maketitle
	
\begin{figure}[th]
		\centering
		\includesvg[width=0.45\textwidth]{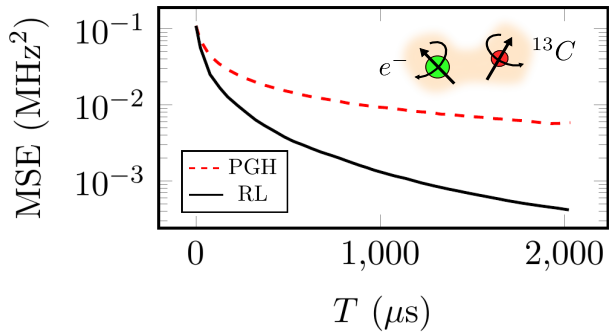}
		\caption{Mean square error (MSE) for the estimation of the parallel hyperfine dipolar coupling between the electron spin and a ${}^{13} C$ nucleus, obtained controlling the MW pulses with the particle guess heuristic (PGH) used in \cite{joas_online_2021}, compared to the performances of model-aware reinforcement learning. The precision is reported as a function of the total measurement time $T$.}
		\label{fig:spinc13}
	\end{figure}

	\textbf{Introduction.} In recent times, there has been a growing focus on the intersection of machine learning and quantum information. The collaboration between these two technological realms holds promise for mutual benefits. Quantum technologies, particularly quantum computers, possess the capability to tackle conventional challenges in machine learning, such as classification and pattern recognition, whether handling classical or quantum data~\cite{broughton_tensorflow_2021, bergholm_pennylane_2022}. Conversely, conventional machine learning can enhance tasks in quantum information, such as quantum control with feedback~\cite{porotti_gradient-ascent_2023} and error correction~\cite{fosel_reinforcement_2018}. Our research falls into the latter category. Specifically, we employ model-aware reinforcement learning to discover optimized adaptive and non-adaptive control strategies for tasks in quantum metrology and estimation. Through this approach, we investigate how machine learning has the potential to improve traditional methods in quantum physics and contribute to the advancement of new quantum information processing technologies. The present article serves as a three-pages extended abstract to the papers containing the theoretical development of this framework~\cite{belliardo_model-aware_2024}, and the applications~\cite{belliardo_applications_2024}. We refer also to the online documentation of the qsensoropt library~\cite{qsensoropt_doc} for details on the implementation and the usage of the framework, and to the repository~\cite{qsensoropt_repo} for accessing the code.
 In a quantum metrology experiment we are interest in the estimation of some unknown parameters and the goodness of the experiment and the data processing can be gauged by an error figure of merit, e.g. the mean square error relative to the true values of the unknown parameters. After specifying a set of adjustable variables within an experiment, an agent trained with reinforcement learning can effectively manipulate them and minimize the error metric. This agent can take the form of a compact neural network, a decision tree, or a straightforward list of trainable controls applied sequentially. The whole controlled estimation has been abstracted from the specific sensor and physical platform and encapsulated into the qsensoropt library, which is accessible on PyPI. Consequently, this library serves as a versatile tool for optimizing a diverse range of quantum sensors. Our framework was tested across various examples using the nitrogen-vacancy (NV) center platform~\cite{chen_quantum_2018, rembold_introduction_2020}, encompassing DC~\cite{fiderer_neural-network_2021} and AC magnetometry, decoherence estimation~\cite{arshad_real-time_2024}, and hyperfine coupling characterization~\cite{joas_online_2021}. We report as example in this work the estimation of the parallel hyperfine coupling of the electron spin in a NV center with a neighbouring ${}^{13} C$ in the carbon lattice, see \cref{fig:spinc13}, in which machine learning produces a better control policy than the adaptive strategy currently used in experiments.
	Additionally, within the realm of photonic circuits, we explored multiphase estimation, a recent extension of the Dolinar receiver~\cite{zoratti_agnostic-dolinar_2021}, and its adaptation to the discrimination of three states, along with coherent states classification. In the frequentist estimation domain, our investigation focused on the sensing of the detuning frequency in a driven optical cavity~\cite{fallani_learning_2022}. Our results demonstrate that model-aware reinforcement learning surpasses traditional control strategies across multiple scenarios. This research lays the groundwork for accelerating the quest for optimal controls in quantum sensors, potentially accelerating their widespread industrial application.
	\begin{figure}[th]
		\centering
		\includesvg[width=0.45\textwidth]{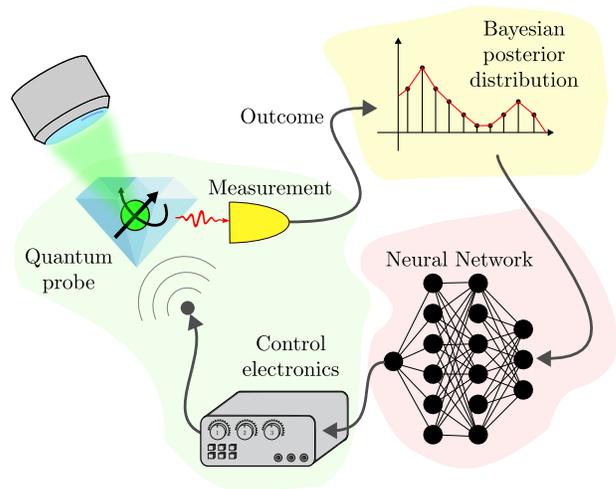}
		\caption{
			This general scheme illustrates the information flow within the measurement loop. The environment we aim to study interacts with the quantum probe and encodes it with the unknown variables $\vtheta$. This probe is then measured using a tunable instrument. The outcome of this measurement provides us with information about the probe's state and in turn about the environment's variables. This information is used in the particle filter to update the posterior Bayesian distribution on $\vtheta$. Some summary information derived from the particle filter is then input into an agent that decides the new control parameters for the measurement in the next iteration of the loop. This control is then realizes through the electronics of the experiment. In this picture, the agent is a neural network.}\label{fig:pipeline}
	\end{figure}

	\textbf{The measurement loop.} In quantum metrology and estimation we analyse a physical system, called quantum probe, governed by a well-known quantum dynamic, which the experimenter can continuously modify by selecting values for a fixed set of controls (e.g. a tunable phase in an interferometer). The experimenter is interested in estimating certain unknown parameters, denoted by $\vtheta$, associated with the environment and encoded in the probe through its interaction, or associated to the initial state of the probe itself. These parameters are estimated through measurements on the probe. The estimation process employs a particle filter~\cite{del_moral_nonlinear_1997, arulampalam_tutorial_2002, liu_sequential_1998} (PF) to process the outcomes of the measurements, which uses the Bayes' rule to update the Bayesian posterior probability distribution on the parameters $\vtheta$ after each measurement. The PF contains a set of discrete samples from the theoretical posterior distribution, named particles, with a weight associated to each of them, to represent the posterior probability distribution. Controls are determined based on information within the PF, such as the mean and variance of the distribution, which are the input to the agent that produces the controls as output. The next measurement is then performed and the process is repeated to form the \textit{measurement loop}, of which a single iteration is represented in \cref{fig:pipeline}. The knowledge of the parameters $\vtheta$ in the PF obtained through measurements is leveraged to guide the evolution and measurements on the probe, optimizing the overall performance of the estimation task.

	\textbf{The sensors model.} While the Bayesian filtering, the probe's control, and the agent's training have been implemented in the library, users must implement a differentiable model of the sensor of their interests using TensorFlow. This model should simulate the stochastic extraction of measurement outcomes and evaluate the probability of observing a specific outcome in a measurement. The model-aware reinforcement learning approach enables us to address a diverse range of tasks using a unified tool.

	\textbf{The precision resource-paradigm.} Every iteration of the measurement loop consumes some amount of a specific ``resource'', which is costly in the context of the estimation and must be defined by the user. Once these resources are depleted, the measurement loop concludes. Examples of resources are the total estimation time, the number of measurement performed or the intensity of a signal, which is continuously probed. 
	
	\textbf{Training loss.} Once the estimation ends and the measurement loop is terminated an estimator $\hvtheta$ for the parameters $\vtheta$ is computed. From this, the user-defined precision metric for the sensor is evaluated. This might be, for instance, the mean square error in a parameter estimation task or the error probability if there is a finite set of possible values for $\vtheta$. This precision metric is the loss to be minimized in the training.
	
	\textbf{Training of the agent.} The training of the agent for each experiment is facilitated by the functions of our library. For the majority of the applications we opted for neural networks due to their proven suitability to approximate generic smooth functions~\cite{de_ryck_approximation_2021}. Through automatic differentiation over all iterations of the loop, a gradient descent procedure is employed to train the agent to minimize the loss. With repeated application of the rule for differentiating composite functions the gradient is computed through the stochastic measurement outcome extraction, the update of the probe state, and the Bayes' rule applied to the PF. By default, the gradient doesn't propagate through the agent's input. Allowing the derivatives to traverse the physical model implementation of the sensor categorizes this training as a form of model-aware policy gradient reinforcement learning. Since the loss is a stochastic variable, as it depends on the simulated measurement outcomes, special precautions are necessary to compute an unbiased estimator for its gradient, which have been firstly introduced in the procedure of feedback GRAPE~\cite{porotti_gradient-ascent_2023} in the context of physics.
	
	\textbf{Conclusions.} In summary, our research underscores the advantages of integrating machine learning with contemporary quantum technologies. We have introduced a framework, complemented by a versatile library, designed to address a broad range of challenges in quantum sensing. This library provides a flexible interface, allowing researchers to easily configure and optimize diverse parameter estimation tasks based on quantum systems. With the potential to expedite the development of practical applications in quantum parameter estimation and metrology, our library opens avenues for precise estimation of physical parameters that could transform various sectors, including biology, fundamental physics, and quantum communication. By offering a user-friendly tool, we aim to facilitate progress in these domains, facilitating the transition of quantum-based metrology from proof-of-principle experiments to industrial applications.
	
	\textbf{Acknowledgments.} We gratefully acknowledge the computational resources of the Center for High Performance Computing (CHPC) at SNS. We acknowledge financial support by MUR (Ministero dell'Istruzione, dell'Università e della Ricerca) through the following projects: PNRR MUR project PE0000023-NQSTI, PRIN 2017 Taming complexity via Quantum Strategies: a Hybrid Integrated Photonic approach (QU-SHIP) Id. 2017SRN-BRK.

	\bibliographystyle{naturemag}
	\bibliography{bibliografia.bib}
	
\end{document}